\documentclass[aps,floats,prl,twocolumn]{revtex4}

\usepackage{graphicx}
\usepackage{docs}

\def\be{\begin{equation}}
\def\ee{\end{equation}}
\def\ba#1{\begin{array}{#1}}
\def\ea{\end{array}}
\def\bn{\begin{enumerate}}
\def\en{\end{enumerate}}

\def\rr{\right}
\def\l{\left}

\def\tpsi{\tilde{\psi}}

\def\summ{\sum\limits}

\def\LL{{\cal L}}

\begin{document}

\title{Sagnac interference in Carbon nanotube loops}
\author{{\sc Gil Refael$^1$, Jinseong Heo$^2$, Marc Bockrath$^2$}\\
{$^1$\small \em Dept. of Physics, California Institute of Technology, MC 114-36, Pasadena, CA 91125}\\
{$^2$ \small \em Dept. of Applied Physics, California Institute of Technology, MC 114-36, Pasadena, CA 91125}}

% -------------------------%
\begin{abstract}

In this paper we study electron interference in nanotube loops. The conductance as a function of the applied voltage 
is shown to oscillate due to interference between electron beams
traversing the loop in two opposite directions, with slightly different
velocities. The period of these oscillations with respect to the gate
voltage, as well as the temperatures required for the effect to
appear, are shown to be much larger than those of the related Fabry-Perot
interference. This effect is analogous to the Sagnac effect in light interferometers. 
We calculate the effect of interactions on the period of the oscillations, and show that even
though interactions destroy much of the near-degeneracy of velocities in the
symmetric spin channel, the slow interference effects survive.  
\end{abstract}
\pacs{PACS Numbers:}

\maketitle
%\maketitle

%------------------ Introduction

Transport measurements on single-walled carbon nanotubes have provided many clear demonstrations of quantum strongly 
correlated phenomena in mesoscopic physics \cite{Devoret}. Particularly exciting 
examples include Luttinger-liquid behavior \cite{KaneBalentsFisher,BockrathBalents}, and the Fabry-Perot interference \cite{Bockrath, PecaBalents} of electrons. 
These effects, in principle, allow the determination of the interaction 
parameters of the Luttinger liquid. Nevertheless, previous calculations 
showed only interference between spin and charge modes with energy 
scales that were of the same order of magnitude, making experimental 
observation challenging. In this paper, we propose, analyze, and show
initial observations of a new mode of interference - Sagnac
interference between two time-reversed paths \cite{Sagnac}- in nanotubes forming a loop (i.e. a nanoloop), and with 
Fermi-surface away from the particle-hole symmetric points. This
effect produces large-period fluctuations of the conductance as a
function of gate-voltage, and source-drain voltage. 
Due to its large expected period, this fluctuation effect is expected to survive
to very high temperatures.

\begin{figure}
\includegraphics[width=8cm]{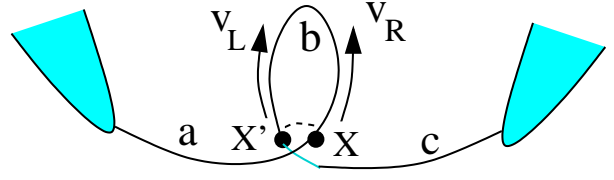}
\caption{The system of interest is a nanotube that forms a loop. At
  the basis of the loop electrons can tunnel (dashed line) from one branch (X) to
  the other (X'). This produces interference between
  counter-propagating electrons due to the velocity- difference between right and left-moving electrons. 
 \label{nanoknot}}
\end{figure}

\begin{figure}
\includegraphics[width=9cm]{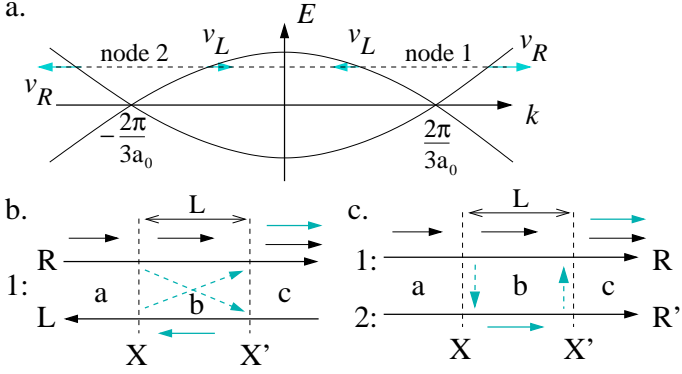}
\caption{{\bf (a)} Dispersion of an armchair nanotube. When the gate voltage
  is nonzero, the right and left movers in each node have different
  velocities, $v_{R/L}=v_F\pm u$ ($a_0$ is the lattice constant). A nonzero $u=(v_R-v_L)/2$
  leads to the two Sagnac interference effects: {\bf (b)} Within one
  node (say node 1), a beam
entering the loop from the left (short black arrow) splits by partially tunneling between points X and X'
to two counter-propagating beams (black and gray), in region b. They then recombine at point
X'. Long black arrows
  represent the two chiral electronic modes near node 1. The regions a, b, and c correspond
  to those indicated in Fig. \ref{nanoknot}. {\bf (c)} A beam impinging on point
  X in node 1 (2) partially
scatters to node 2 (1). Both traverse the loop in the same
direction (short black and gray arrows), and recombine at point
  X'. This effect is similar to the slow conductance oscillation 
due to impurities propounded in
  Ref. \onlinecite{Jiang2003}, and in the presence of axial magnetic field in
  Ref. \onlinecite{AxialField}. 
\label{interference}}
\end{figure}

The Sagnac interferometer measures the angular velocity of a ring, by
measuring interference fringes between light propagating inside the
ring in two opposite directions. Here we consider the Sagnac effect in an armchair nanotube with a loop (Fig. \ref{nanoknot}), i.e., two points
along the nanotube between which electrons can tunnel (such tunneling
effects were discussed in Ref. \onlinecite{KomnikEgger,PotsmaDekker}). 
Electrons impinging on point $X$ in Fig. \ref{nanoknot}, can either
proceed and traverse the loop in a counter-clock-wise direction, or
tunnel to point $X'$, and traverse the loop in a clock-wise
direction, reproducing the Sagnac interference effect (Fig. \ref{interference}b). Note that unlike Fabry-Perot interference, in which the interfering beams
traverse a different physical distance (e.g. the loop a
different number of times) the Sagnac interference is between two
beams traversing {\it exactly} the same distance. 

The role of the
angular velocity for the light interferometer is replaced with a
velocity difference for the two counter propogating electronic
beams (Fig. \ref{interference}b). Using a gate voltage $V_g$, the Fermi-surface is tuned away from the band middle; ignoring interactions, 
right and left moving electrons in one node may have different
velocities: $v_R=v_F+u$, and $v_L=v_F-u$ (Fig. \ref{interference}a)\cite{disp}. 
A small velocity difference $u$, like a small angular velocity in
the light Sagnac effect \cite{Sagnac}, produces a slow fluctuation of the conductance as a
function of $V_g$ \cite{foot1}. The phase difference between the two interfering beams is:
\be
\Delta \phi=L k_L-L k_R=\frac{L \epsilon_F}{\hbar
  v_L}-\frac{L\epsilon_F}{\hbar v_R}\approx \alpha e V_g
L\frac{2u}{\hbar   v_F^2}
\label{dphi}
\ee
where $L$ is the length of the loop, and $\epsilon_F=\hbar v_{R/L}
k_{R/L}$. Also, $\epsilon_F=\alpha e V_g$, where $\alpha$
is the conversion factor between the gate voltage and change in
chemical potential. Interference fringes repeat when
$\Delta\phi=2\pi n$. Since roughly $u\propto V_g$, the n'th
fringe is at $V_g\propto \sqrt{n}$; fringes are more dense as we move away from the middle of the nanotube's
band. For non-interacting electrons,
the same fringes should appear as a function of a source-drain
voltage, $V_{sd}$. 
In the armchair-tube nanoloop, beams moving in the same direction
around the loop, but in different
nodes, also interfere (Fig. \ref{interference}c). The two beams in this {\it band-Sagnac} effect differ by the same phase due to the time-reversal symmetry.

%-------------- description of the nanoknot

%-------------------- interactions 

The simple analysis above, which ignores interactions, already provides a good picture of the
Sagnac effect in nanotubes. Nevertheless, thin single-walled nanotubes are expected to have a Luttinger parameter $g\sim
0.3$ \cite{Bockrath}. Interactions change the hydrodynamic
velocities in the nanotube dramatically, and may lift the near
degenracy of the velocities between the interfering beams. In the
following we analyze the Sagnac interference effect (and also the
Fabry-Perot interference with $u\neq 0$) of interacting electrons. We will show
that interactions do not destroy the large-period Sagnac
fringes. Yet strong modifications exist: the fringes in the
conductance as a function of $V_g$ are determined mostly by the bare, non-interacting,
velocity spectrum, essentially reflecting Eq. (\ref{dphi}),  while the fringes in $V_{sd}$ are modified dramatically, and
reflect the four velocities of the tube's hydrodynamic modes.

The bosonized Lagrangian of the two Dirac nodes, with $\lambda$ parametrizing the density-density
interaction, is:
\be
\ba{c}
\LL=\frac{\hbar v_F}{2\pi}\summ_{\sigma,a=1,2}\int dx \l[\frac{2}{v_F}\dot{\theta}_a^{\sigma}\nabla
\phi_a^{\sigma}-\l(\nabla\theta_a^{\sigma}\rr)^2-\l(\nabla\phi_a^{\sigma}\rr)^2\rr.\\\l.+(-1)^a 2\frac{u}{v_F}\nabla\phi_a^{\sigma}\nabla\theta_a^{\sigma}\rr]
-\int dx \lambda\l(\summ_{\sigma,a=1,\,2}\frac{1}{\pi}\nabla\theta^{\sigma}_a\rr)^2.
\ea
\label{lag}
\ee
The subscript $a$ designates the Dirac node, $\sigma$ is the
spin. The charge mode, $\frac{1}{2}\summ_{\sigma,\,a=1,2}\l(\theta^{\sigma}_a\rr)$,
has the interaction Luttinger parameter $g=(1+8\lambda/\pi \hbar v_F)^{-1/2}$.  The
Fermionic operators are: $\psi^{\sigma}_{a\,R/L}\sim e^{i
  (-1)^a  \frac{2\pi}{3a_0}x-i\l(\phi^{\sigma}_a\pm\theta^{\sigma}_a\rr)}$.

\begin{figure}
\includegraphics[width=7cm]{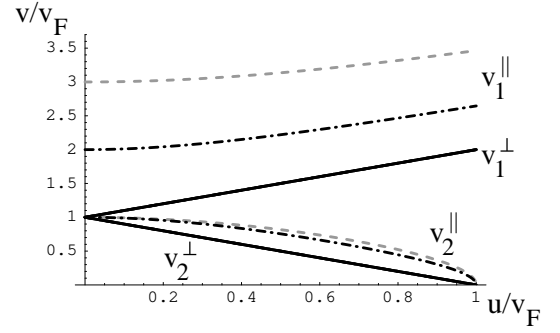}
\caption{The velocities of the four hydrodynamic modes as a function of
  left-right asymmetry of the non-interacting spectrum. Roughly
  speaking, $v^{||}_1$ corresponds to the charged mode (spin and node
  symmetric), and $v^{||}_2$ is spin symmetric but node
  antisymmetric. $v^{\perp}_{1,\,2}$ are due to the spin-anti-symmetric
  modes in nodes 1 and 2. The solid line has $g=1$, dashed-dotted has $g=1/2$,
  and dashed gray has $g=1/3$. \label{velocities}
}
\end{figure}

The Fourier-transform of (\ref{lag}) defines an
$8\times 8$ quadratic form of the $\theta$'s and $\phi$'s. Its eight
eigenvalues are square-roots of second degree polynomials of $\omega$ and $k$. The dispersion of the
eight chiral modes is given by the values of
$\omega/k$ which make an eigenvalue vanish. We find
that the spin anti-symmetric channel consists of four untouched neutral chiral modes, with
velocities:
\be
\ba{cc}
v^{\perp}_{1\,\pm}=\pm v_F+u &
v^{\perp}_{2\,\pm}=\pm v_F-u
\label{Ivel}
\ea
\ee
where $+$ and $-$ indicate right and left movers respectively. More
interestingly, the remaining four spin-symmetric chiral modes are
given by (see also Ref. \onlinecite{Oreg95}):
\be
\frac{v^{||}_{1,\,2}}{v_F}=\frac{1}{\sqrt{2}}\sqrt{1+\frac{1}{g^2}+2\frac{u^2}{v_F^2}\pm\sqrt{\l(1-\frac{1}{g^2}\rr)^2+8\frac{u^2}{v_F^2}\l(1+\frac{1}{g^2}\rr)}}
\label{gvel}
\ee
$v^{||}_{1}$ and $v^{||}_{2}$ describe two right-left symmetric
branches of the spectrum. $v^{||}_1$ is the charge mode velocity, when $u=0$. The four velocities are depicted in
Fig. \ref{velocities}. In the absence of spin scattering, only the
spin-symmetric modes can interfere with each other, but the
velocities related to these two modes are very different, as can be
seen by Eq. (\ref{gvel}). In the following, we determine which velocities appear, then, in the
interference fringes of $V_g$ and $V_{sd}$.

$V_g$ couples to the total
electronic density, whereas $V_{sd}$ couples to the density difference of
right and left movers:
\be
\LL_{g+sd}=\int dx \l(\alpha e V_g \frac{1}{\pi}\nabla\summ_{\sigma,\,a}\theta^{\sigma}_a+e V_{sd} \frac{1}{\pi}\nabla \summ_{\sigma,\,a}\phi^{\sigma}_a\rr). 
\label{Vg}
\ee
$V_g$ changes the chemical potential and Fermi-surface of the
electrons. This is seen by absorbing the new term in the $\theta$ and
$\phi$ gradient terms (where the latter is involved only due to
the difference in the original right and left moving
velocities). Unlike $V_g$, $V_{sd}$ drives the system out of equilibrium: it induces a
current. This entails a time-dependent transformation of the
bosonic fields to absorb the term. 
The transformation:
\be
\ba{cc}
\tilde{\theta}^{\sigma}_a=\theta^{\sigma}_a-\frac{\alpha g^2 e V_g}{\hbar
  v_F}\frac{1}{1-g^2u^2/v_F^2} x+\frac{eV_{sd}}{\hbar} t,\\
\tilde{\phi}^{\sigma}_a=\phi^{\sigma}_a-(-1)^a \frac{\alpha g^2 e V_g}{\hbar
  v_F}\frac{u/v_F}{1-g^2u^2/v_F^2} x,
\ea
\label{gatetrans}
\ee
absorbs both $V_g$ and $V_{sd}$ in the bosonic fields, with $\sigma=\uparrow,\,\downarrow$, and $a=1,\,2$
is the node. The slow, $u$-dependent, fluctuations
can already be noticed in $\phi^{\sigma}_a$'s $x$-dependence.
The above procedure is drawn from Ref. \onlinecite{PecaBalents}. 

Following are the possible scattering processes contributing
to transport through the nanoloop. For simplicity, we define 
$\tilde{\psi}^{\sigma}_{a\,R/L}\sim e^{i (-1)^a \frac{2\pi}{3a_0}x-i\l(\tilde{\phi}^{\sigma}_a\pm\tilde{\theta}^{\sigma}_a\rr)}$.
The simplest term is the {\bf same-node
back-scattering}:
\be
\hat{B}(x,t)=b\summ_{\sigma,\,a}\l(\tilde{\psi}^{\sigma\dagger}_{a~R}(x)\tilde{\psi}^{\sigma}_{a~L}(x)e^{2i k_g x-2i\omega_{sd} t}+\mbox{h.c.}\rr),
\label{Back}
\ee
\be
\ba{lcc}
\mbox{with:}\hspace{0.5cm} & k_g=\frac{\alpha e g^2 V_G}{\hbar v_F\l(1-g^2u^2/v_F^2\rr)}, &
\omega_{sd}=eV_{sd}/\hbar v_F.
\ea
\label{kgg}
\ee
Second is {\bf cross-node back-scattering}: $\hat{N}_b(x,t)=$
\be
n_b\summ_{\sigma,\,a}\l(\tpsi^{\sigma\dagger}_{a~R}(x)\tpsi^{\sigma}_{\overline{a}~L}(x)
e^{2ik_g\l(1+(-1)^a \frac{u}{v_F}\rr)x-2i\omega_{sd} t}+\mbox{h.c.}\rr),
\label{CBBack}
\ee
which is backscattering from node $a$, to node, $\overline{a}=3-a$.  
Third is {\bf cross-node forward-scattering}: $\hat{N}_f(x,t)=$
\be
\ba{c}
n_f\summ_{\sigma,\,a}\l(\tpsi^{\sigma\dagger}_{a~R}(x)\tpsi^{\sigma}_{\overline{a}~R}(x)+\tpsi^{\sigma\dagger}_{a~L}(x)\tpsi^{\sigma}_{\overline{a}~L}(x)\rr) e^{2i(-1)^a\frac{k_g u}{v_F} x}.
\label{CBFor}
\ea
\ee
Most important is the backscattering term arising from
tunneling between point X at $x=0$,
and X' at $x=L$ (Fig. \ref{interference}a) , i.e., {\bf cross-loop back-scattering}:
\be
\ba{c}
\hat{K}_b(t)=k_b\summ_{\sigma,a,b}
\l(\tpsi^{\sigma\dagger}_{a~R}(0)\tpsi^{\sigma}_{b~L}(L)e^{ik_g\l(1-(-1)^b\frac{u}{v_F}\rr) L-2i\omega_{sd}t}\rr.\\\l.
+\tpsi^{\sigma\dagger}_{a~L}(0)\tpsi^{\sigma}_{b~R}(L)e^{-ik_g\l(1+(-1)^b\frac{u}{v_F}\rr) L+2i\omega_{sd}t}
+\mbox{h.c.}\rr).
\label{CLback}
\ea
\ee

To calculate the conductance fluctuations,
we must follow the Kubo/Keldysh formalism as it applies to the various
scattering events, $\hat{L}_{m}(x,t)$, where $\hat{L}_m=\hat{B},\,\hat{N}_{f/b}, \hat{K}_b$. We defer an exact
evaluation to a later publication, and concentrate here on the main
features of the fluctuations:
\be
\Delta G_{mn}\sim \int_0^{\infty} dt
\langle\l[\hat{L}_m(t),\hat{L}_n(0)\rr]\rangle\sim\langle\hat{L}_m\hat{L}_n\rangle_{\omega=2eV_{sd}/\hbar}.
\label{Kubo}
\ee
The second relation, connecting the integral to the
correlation's Fourier transform, is due to the time dependence of
the integrand being $e^{2i\omega_{sd}t}=e^{2ieV_{sd}t/\hbar}$. In the case of $\hat{L}_m$ occurring at $x=0$ and $\hat{L}_m$ at $x=L$,
the dependence of the oscillating part of $\Delta G_{mn}$ on $V_{sd}$ is easily seen to be of
the form:
\be
\Delta
G_{mn}=f\l(e^{2i\omega_{sd}L/v^{||}_1},e^{2i\omega_{sd}L/v^{||}_2},e^{2i\omega_{sd}L/v^{\perp}_{1+}},e^{2i\omega_{sd}L/v^{\perp}_{2+}}\rr),
\label{sdint}
\ee
where $v^{||/\perp}_{1/2}$ are the four propagation velocities
of the various hydrodynamic modes found in Eqs. (\ref{gvel}) and
(\ref{Ivel}). Thus we find that the interference fringes as a function
of $V_{sd}$ are determined by the velocities of the
interaction-induced four hydrodynamic modes. 

Our main result is the interference pattern in $G$
vs. $V_g$.  From the various
scattering events we infer that only two expressions give rise to
interference effects in $V_g$: $\exp(2ik_g L)$ and
$\exp(2ik_g  L\frac{u}{v_F})$, with $k_g$ defined in
Eq. (\ref{kgg}). Table \ref{Gtab} enumerates the conductance fluctuations due
to second-order scattering. Our focus is the loop-Sagnac interference between
counter-propagating beams in an interacting nanotube, which is the
first line in Table \ref{Gtab}). Indeed, it coincides with the
band-Sagnac interference (B-SAG); it is possible, however, to distinguish the two Sagnac
modes by applying a magnetic flux to the loop. The band-Sagnac fringes will be unaffected, whereas the
loop-Sagnac phase-difference will be shifted by $2e\Phi/\hbar$, where
$\Phi$ is the flux through the loop. 

\begin{table}
\begin{center}
\begin{tabular}{|c|c|c|c|}
\hline
type & $\hat{L}_m\hat{L}_n$ & $\delta V_g$ & $T_c$ \\
\hline
L-SAG & $\hat{K}_b\hat{K}_b$ &
$\frac{\pi\hbar
  v_F}{\alpha e g^2 L}\l(1-\frac{u^2 g^2}{v_F^2}\rr)\frac{v_F}{u}$ &   $\frac{\pi\hbar
  v_F}{L}\l(1-\frac{u^2 g^2}{v_F^2}\rr)\frac{v_F}{u}$\\
B-SAG &  $\hat{N}_{f}\l(\hat{N}_{b}\hat{B}\rr)$ &
$\frac{\pi\hbar
  v_F}{\alpha e g^2 L}\l(1-\frac{u^2 g^2}{v_F^2}\rr)\frac{v_F}{u}$ &  $\frac{\pi\hbar
  v_F}{L}\l(1-\frac{u^2 g^2}{v_F^2}\rr)\frac{v_F}{u}$\\
FP & $\hat{B}\hat{B}$ & $\frac{\pi\hbar
  v_F}{\alpha e g^2 L}\l(1-\frac{u^2 g^2}{v_F^2}\rr)$ & $\frac{\pi\hbar
  v_F}{L}\l(1-\frac{u^2 g^2}{v_F^2}\rr)$\\
B-FP$\pm$ & $\hat{N}_b\hat{N}_b$ &  $\frac{\pi\hbar
  v_F^2}{\alpha e g^2 L}\frac{1-u^2 g^2/v_F^2}{v_F\pm u}$ &  $\frac{\pi\hbar
  v_F^2}{L}\frac{1-u^2 g^2/v_F^2}{v_F\pm u}$\\
L-FP$\pm$ & $\hat{K}_b\hat{B},\,\hat{K}_b\hat{N}_b$ & $\frac{2\pi\hbar
  v_F^2}{\alpha e g^2 L}\frac{1-u^2 g^2/v_F^2}{v_F\pm u}$ & $\frac{2\pi\hbar
  v_F^2}{L}\frac{1-u^2 g^2/v_F^2}{v_F\pm u}$\\
\hline
\end{tabular}
\caption{Interfering contributions to the conductance as a function
  of $V_g$. The first row indicates the type of
  interference. The loop-Sagnac (L-SAG) and band-Sagnac (B-SAG)
  correspond to Fig. \ref{interference}b,c respectively.
The three Fabry-Perot modes originate from optical-path difference:
  of two-loops in node 1 {\it or} 2 (B-FP$\pm$), of one loop in node 1 and
  one in 2 (FP), or, due to loop-tunneling, one loop in node 1 {\it or} 2 (L-FP$\pm$).
The coherence temperature of each interference mode is
  determined heuristically by assuming that $T_c\sim \alpha g^2 e
  \delta V_g/k_B$.\label{Gtab}}
\end{center}
\end{table}

A remarkable difference between the Sagnac and Fabry-Perot interference is the sensitivity to temperature. Using a simple
argument we estimate the maximum temperature for each interference. The kinetic
energy of a single electron has uncertainty of order $T$. This could
be thought of as an uncertainty in the gate voltage:  $\Delta
V_g\sim T_c/\alpha g^2 e$. When $\Delta V_g\approx \delta V_g$, an
interference fringe disappears, which is how we obtain $T_c$ in Table
\ref{Gtab}. A subtle
point is the absence of $g^2$ in the $T_c$ column in Table \ref{Gtab}; since temperature
only smears the kinetic energy of electrons, the effects of
interactions should be omitted to first approximation, hence the
cancellation of $\alpha g^2$. We find that $T_c$ for the Sagnac modes is $v_F/u$ larger than that of the Fabry-Perot interference. 
\be
T_{c}^{SAG}\sim \frac{\pi\hbar v_F}{L}\l(1-\frac{u^2
  g^2}{v_F}\rr)\frac{v_F}{u}\sim \frac{v_F}{u} T_c^{FP}
\ee
Hence the limiting factor for the observation of the Sagnac effect
is most likely phonon scattering.  

\begin{figure}
\includegraphics[width=7cm]{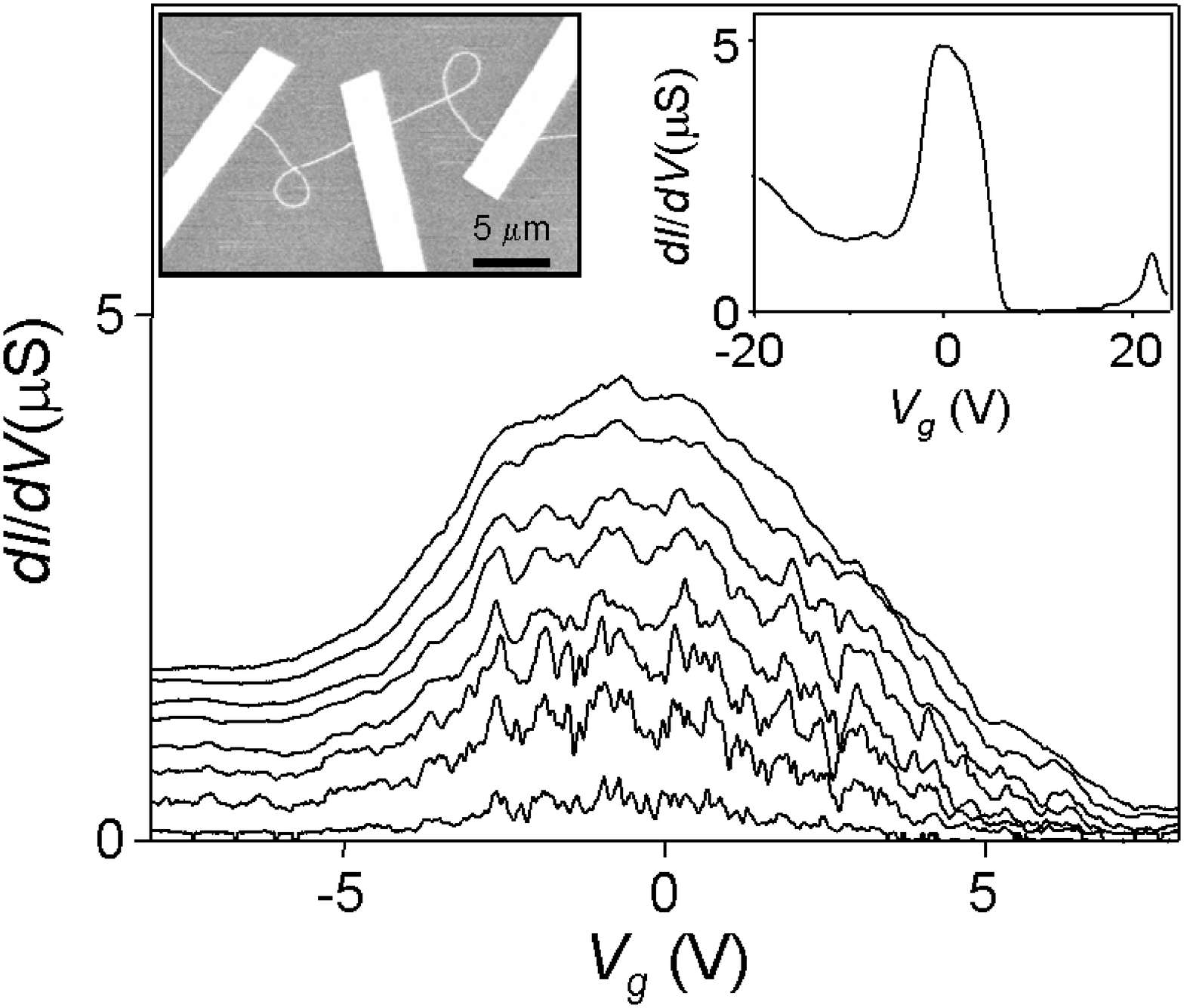}
\caption{Conductance vs. gate-voltage of a nanoloop
  device. From top to bottom:
  $T=64K,\,48K,\,32K,\,24K,\,16K,\,12K,\,8K,\,4K$. Strong conductance fluctuations appear already at $T=64K$
  with period $\delta V_G\sim 20V$. These fluctuations are consistent with
  the Sagnac interference effects. At lower
  temperatures higher-frequency Fabry-Perot interference appears as
  well with periods $\delta V_g\sim 0.15V,\,0.3V$. Left inset shows a
  scan of the device; data from both loops is qualitatively the same. Right inset shows the suspected Sagnac envelope
  at T=4K and $V_{sd}=30mV$ for wider range of $V_g$. Note that the asymmetry in $V_g$ is not
  understood. \label{expfig}} 
\end{figure}

The experimental observation motivating this work is shown in
Fig. \ref{expfig}. At $T=32K$ fast oscillations with period $\delta
V_g\sim 0.3 V$ appear; we identify them with the loop-FP mode of
Table \ref{Gtab} (Coulomb blockade is determined by the total wire
length, and is expected at a much
lower $\delta V_g\sim 10^{-3}-10^{-2}V$). Another fast mode appears at $T=12K$, with a
doubled frequency, $\delta V_g\sim 0.15 V$, and therefore fits
the regular FP mode. But in addition, a slowly oscillating envelope of the
conductance is already evident at $T=64K$, with the first period being roughly
$\delta V_g\sim 20V$\cite{foot2}. If we identify this with the Sagnac
effects, then $\delta V_g^{SAG}/\delta V_g^{FP}\sim 130$.
To see if this is indeed feasible, we approximate the
nanotube's dispersion as
parabolic,  $\pm\epsilon_k=-\gamma(1-\l(\frac{2\pi}{3a_0}\rr)^2 k^2)$;
and $\frac{u}{v_F}\approx \frac{\epsilon}{2\gamma}$. The first fringe due
to the Sagnac interference appears when the
accumulated phase difference between the two interfering beams is
$\pi$:
\be
\Delta\phi=\int_0^{\mu^{SAG}} d\epsilon \frac{L}{\hbar
  v_F}\frac{2u}{v_F}\approx \frac{L}{\hbar
  v_F}\frac{(\mu^{SAG})^{2}}{2\gamma}=\pi,
\ee 
where the integral is necessary due to the dependence of $u/v_F$ on
$\epsilon$. The first fringe due to Fabry-Perot
interference is when $\mu^{FP} 2L/\hbar v_F=\pi$. Using $L=7\mu m$ and $v_F=8\cdot 10^5
m/s$, and $\gamma\approx 2.5 eV$ \cite{disp}, we obtain
$V_g^{SAG}/V_g^{FP}=\mu^{SAG}/\mu^{FP}\sim 300$, which agrees with the
experiment up to a factor of 2. This extra factor might be due to the Fabry-Perot
interference arising not from the loop, but from the shorter sections
of the nanotube. The experimental results are also consistent with the fact that the
Fabry-Perot interference are expected roughly at: $T_c^{FP}\sim
\frac{2\pi\hbar v_F}{L}\sim 10K$. Since
currently only two samples of the loop geometry are
available, we limit ourselves to the order-of-magnitude analysis
above, and defer a detailed analysis of the experiment to a future publication. 

In addition to the Sagnac interference, we also obtained the
modification of the Fabry-Perot interference in the presence of
right-left asymmetry and interactions. Most interesting in this
respect are the loop- and band- FP effects. In a non-interacting
system, these arise from one of the interfering beams going through
the loop once, or twice, more than the other, but in a single band. These effects, with a minus sign in Table
\ref{Gtab}, $\delta V_g=\frac{\pi\hbar  v_F^2}{\alpha e g^2
L}\frac{1-u^2 g^2/v_F^2}{v_F-u}$, present the {\it only} interference that may increase their
period as $V_g$ tunes away from the Dirac nodes. Slow nanotube
conductance oscillations were also seen in Ref. \onlinecite{Laughlin},
and at first interpreted as Fabry-Perot interference between two
closely-spaced localized impurities on the nanotube.
Ref. \onlinecite{Jiang2003}, tried to explain them as an impurity interference effect, similar to what we call the band-Sagnac effect (Fig. \ref{interference}c). But the period of these
oscillations increases with distance from the middle of the
band, contrary to Ref. \onlinecite{Laughlin} observations. Thus we
conclude that the slow interference of Ref.\cite{Laughlin} may indeed
be due to band-Fabry-Perot interference 
between closely-spaced impurities.

In this paper we discussed the Sagnac and Fabry-Perot interference effects in interacting
nanotube loops, with $v_R-v_L=2u\neq 0$. We found that
$V_g$ changes the 'carrier wave' and induces fluctuations that
depend mostly on the bare dispersion of the nanotube, while $V_{sd}$ produces
fluctuations whose periods depend on the velocities of the
non-equilibrium hydrodynamic modes. By studying these conductance
fluctuations experimentally, one could in principle extract all
hydrodynamic velocities, the interaction parameter, and
the bare electron dispersion. We also provided rough estimates of the coherence
temperatures, $T_c$, required to see the Sagnac interference, and
showed that it is much higher than that of the Fabry-Perot interference. The estimates of
$T_c$ are expected to be modified by a precise inclusion of
interactions; this we will pursue in a future publication, in addition
to the explicit dependence of the conductivity on $V_{sd}$. 
The Sagnac interference is closely related to the interference giving rise to weak localization, therefore its 
analysis could directly determine the temperature and interaction
dependence of the electronic dephasing time $\tau_{\phi}$ (see further
Ref. \cite{Altshuler, vondelft}). Here we showed that the Sagnac
effect clearly survive interactions at $T=0$, therefore our results can
be interpreted as evidence for the divergence of $\tau_{\phi}$ in
an interacting electronic system. 

We are grateful to  J. von Delft and Y. Oreg for a critical
discussion of the manuscript, and to  L. Balents, E. Demler,
D. Feldman, V. Galitski,
and J. Meyer for helpful comments. GR thanks the KITP
where part of this work took place. MB and JH are grateful to
support by the ONR.

\bibliography{nanoloop}

\begin{thebibliography}{18}
\bibitem[{foo({\natexlab{a}})}]{foot1}
\bibinfo{note}{Note that another interesting work that relies on $u\neq 0$ is
  Ref. \onlinecite{Feldman}.}
\bibitem[{foo({\natexlab{b}})}]{foot2}
\bibinfo{note}{Another fainter fluctuation with period $\delta V_g\sim 1V$
  seems present at $T=64K$, we are unsure as to its origin.}
\end{thebibliography}

\end{document}